# Astrometric Measurement of the Double-Star System WDS 09168-3407 B 1118

Elijah Ibharalu[1], Miracle Chibuzor Marcel[2],

1. University of Benin: ibharaluelijah@gmail.com

2. Pan African Citizen Science e-Lab, FCT, Abuja

**Abstract**

We present updated measurements for the position angle and separation of the double-star WDS 09168-3407 B 1118, incorporating data from our observation, Gaia Early Data Release 3 mission, and historical records. Our findings indicate a position angle of 341.4° and separation of 3.36". Additionally, we utilized parallax and the ratio of proper motion metric (rPM) values to evaluate the potential binarity of the system. Our analysis suggests that the system may be gravitationally bound.

## 1.0 Introduction

Double stars have been interesting to astronomers for more than two hundred years. These stars are unique due to their closeness to each other in terms of their position in space. This intriguing celestial phenomenon has been a central subject of astrophysical study from the late 18th century until the early 20th century. The Washington Double-Star Catalog of the United States Naval Observatory has the first observations of binary stars, representing a big step forward in our study of these strange pairs of stars (Letchford et al., 2022). The study of double-star systems is very important in astronomy, especially when it comes to learning about massive stellar development. Astrometric traits of binary interactions need to be looked into in great detail because they have a crucial impact on the fates of stars (Gotberg et al., 2018).

In this paper, we present our observational study of the double star system WDS 09168-3407 B 1118, which is found in the constellation of Pyxis with coordinates RA =09h 16m 50.53s and Dec = -34° 06' 39.3". This study explores this system by applying several datasets, which includes historical data from the Washington Double Star Catalog (Mason et al., 2001), as well as recent observational data from the Las Cumbres Observatory and the Gaia EDR3 mission (Gaia Collaboration et al., 2022). The observational records from the Las Cumbres Observatory offer high-resolution data, increasing our understanding of this double-star system. Meanwhile, the Gaia mission greatly improves the accuracy of our astrometric measurements, giving invaluable insights into the celestial dynamics of this specific star system.

Our main focus involved the exact measurement of the position angle and separation between the constituent stars within the B 1118 system. Rigorous inspection of our results involved a thorough comparison with established studies, ensuring the truth and trustworthiness of our findings within the framework of existing astrometric knowledge. This study goes into a comprehensive examination of the B 1118 system, with an emphasis on revealing its inherent nature and definitive classification. We aim to determine the categorical definition of the system, answering the basic question of whether it fits more closely with the characteristics of an optical double or a binary star.

Table 1 shows known information about the stars studied. The data is collected from Gaia Data Release 3 and the Washington Double Star Catalog.

Table 1. The Right Ascension (RA), and Declination (Dec) were retrieved from the Washington Double Star (WDS) catalog, the magnitudes of the primary and secondary stars were taken from the Gaia Gmag database, the parallax and proper motion of the primary and secondary stars were taken from the Gaia catalog, and the rPM was calculated from the proper motions of the system.

|  | Gaia Gmag | Parallax (mas) | Distance (parsec) | Proper Motion (mas/yr) | rPM |
|---|---|---|---|---|---|
| Primary Star | 9.571 | 1.6725 | 597.9 | pmra = −0.374<br>pmdec = 4.642 | 0.033 |
| Secondary Star | 11.137 | 1.6461 | 607.5 | pmra = −0.235<br>pmdec = 4.578 |  |

The proper motion (PM) of the double star system is given in column 5 of Table 1, detailing the values in both right ascension (RA) and declination (Dec). These numerical values served as the base for finding the ratio of proper motions (rPM) metric, a concept proposed by Harshaw (2016) and illustrated by the equations given below. The rPM, quantifying the differences between the proper motions of the primary and secondary stars, is derived by taking the magnitude of their difference and dividing it by the larger magnitude of the two proper motions. The rPM rates the dissimilarity in proper motions related to the magnitude of the larger proper motion. According to the established standards, if the rPM goes below 0.2, the stars are likely a Common Proper Motion (CPM) pair. If the rPM is between 0.2 and 0.6, the stars show Similar Proper Motion (SPM), and if the rPM exceeds 0.6, they display Distinct Proper Motion (DPM). Applying this approach, B 1118, with a rPM measure value of 0.033, is indicative of being a Common Proper Motion (CPM) pair.

$$\text{Resultant} = \sqrt{(R_{pri} - R_{sec})^2 + (D_{pri} - D_{sec})^2}$$

$$\text{Vector} = \sqrt{R^2 + D^2}$$

$$\text{rPM} = \frac{Resultant}{Vector}$$

where $R_{pri}$ & $D_{pri}$ represents the Right Ascension and Declination of the primary star, $R_{sec}$ & $D_{sec}$ represents the Right Ascension and Declination of the secondary star, R & D represent the Right Ascension and Declination of either the primary and secondary star, but whose vector is higher.

## 1.1 Target Selection

We used Stelle Doppie, a tool that taps into the Washington Double Star Catalog (WDS) database, to select the double-star system for our study carefully. To ensure visibility during the course of our research, we picked right ascension (RA) values between 10 and 22 hours. We confirmed this using

Stellarium, a sky simulation software. The declination (Dec) was not limited to any value by us, given the availability of 0.4m LCO telescopes in both hemispheres. Our attention focused on stars observed before 2015, ensuring observable astrometric changes in the system. The primary star's magnitude was limited between 9 and 11, aligning with the observational capabilities of the 0.4m LCO telescopes with a magnitude limit of 20.5. The secondary star magnitude stayed unspecified, with a deliberate selection for a magnitude difference (Δmag) of less than 4, ensuring both stars were visible (Marcel et al., 2024). The stated parameters led to selecting the double-star system WDS 09168-3407 B 1118, with equatorial coordinates of RA = 09h 16m 50.53s and DEC = -34° 06' 39.3". These data were received from Stelle Doppie.

The B 1118 double star system was singled out for examination based on several considerations:

1. It has been examined just seven times, starting from its first observation in 1911 and ending with its most recent inspection in 2010.

2. The fundamental essence of this system is still uncertain.

3. The most recent measurement of the star dates back fourteen years, allowing a large time for probable relative motion between the two stars.

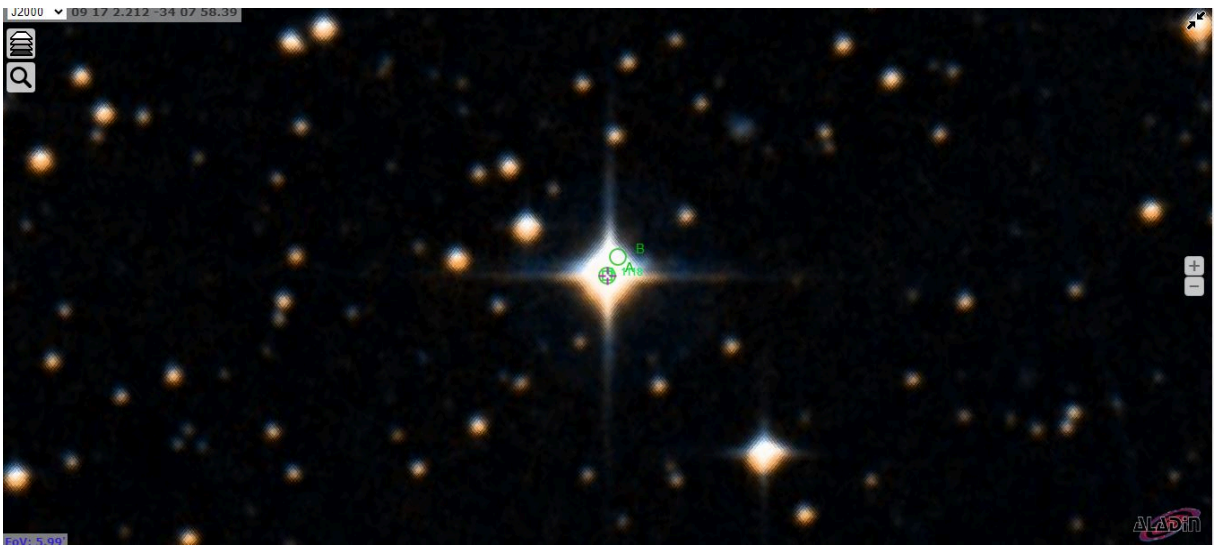

Figure 1: The representation extracted from Stelle Doppie illustrates the target configuration. The dominant hue, indicated in yellow-orange, designates the primary star, while the secondary counterpart is discerned by its comparatively whiter appearance.

## 2.0 Methodology

Observations were performed using Las Cumbres Observatory's Global Telescope (LCOGT), specifically the Cerro Tololo facility in Chile, known for its longstanding role as a main base for astronomy research in the southern skies. Given our star's location in the southern hemisphere, the Cerro Tololo site emerged as the best choice for observations. We used the Bessell-V filter of the 0.4m telescope equipped with an SBIG 6303 CCD camera, known for its field of view measuring 29.2 x 19.5 arcminutes and a plate scale of 0.571 arcseconds per pixel.

The image files were systematically processed by the LCOGT through their automated BANZAI pipeline. Ten pictures were taken, each with an exposure time of 2 seconds. Subsequently, we applied the AstroImageJ program (Collins et al., 2017) to discover the separation and position angle of our star system from the obtained images. This involved choosing an aperture size of 2 pixels within the program. AstroImageJ allowed the automatic determination of the center of the star, utilizing the position as a weighted average of pixel brightnesses within the chosen aperture.

Following the aperture size selection and a closer study of the image area containing the stars, we performed a command drag from the primary to the secondary star to obtain the PA and Sep, (represented as ArcLen). One of our image measurements is illustrated in Figure 3

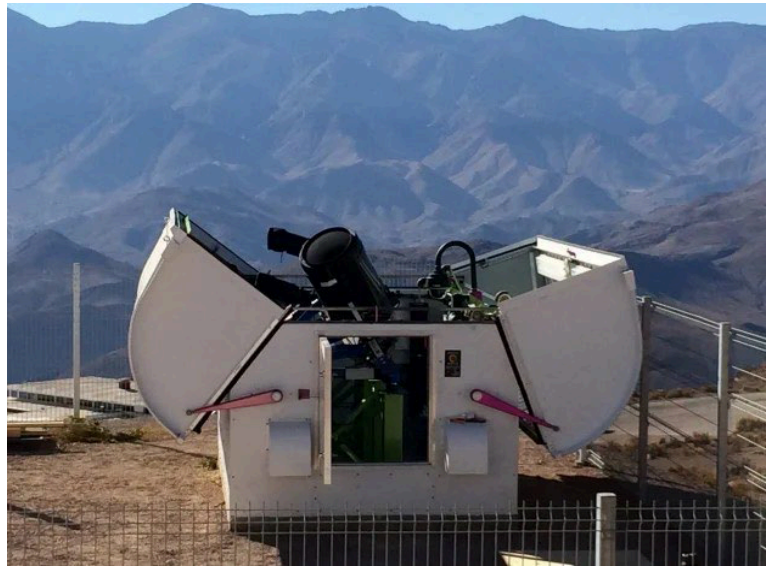

Figure 2: 0.4 m telescope located at one of Las Cumbres Observatory's sites.

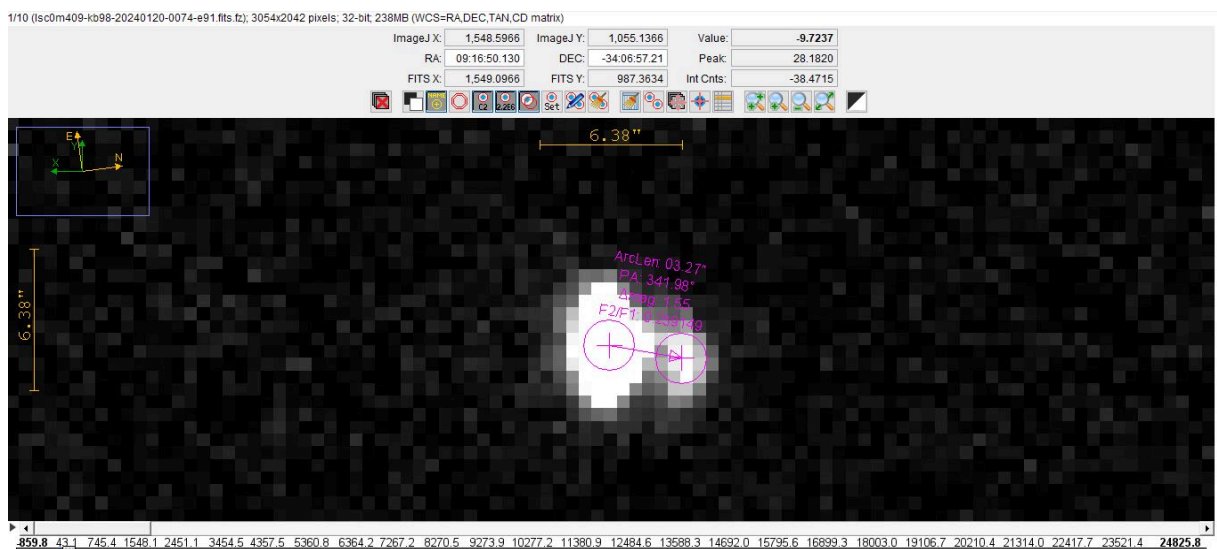

Figure 3: Representative image capturing the WDS 09168-3407 B 1118 double star system sourced from AstroImageJ (AIJ).

## 3.0 Observation

Table 2 presents the newly acquired measurements obtained from our ten images. Table 3 provides a comprehensive summary of statistical analyses conducted on our measurements.

| S/N | PA(°) | SEP(") |
|---|---|---|
| 1 | 342.0 | 3.27 |
| 2 | 340.9 | 3.32 |
| 3 | 343.2 | 3.21 |
| 4 | 341.7 | 3.39 |
| 5 | 341.4 | 3.48 |
| 6 | 343.9 | 3.22 |
| 7 | 339.7 | 3.49 |
| 8 | 340.9 | 3.47 |
| 9 | 341.6 | 3.46 |
| 10 | 338.8 | 3.26 |

Table 2: Novel measurements of WDS 09168-3407 B 1118 derived from our image analysis.(link to [images](images))

| Double Star | Date | Images | | PA(°) | SEP(") |
|---|---|---|---|---|---|
| WDS 09168-3407 B 1118 | 21st January, 2024 (2024.1425) | 10 | Mean | 341.4 | 3.36 |
| | | | Standard Deviation | 1.50 | 0.114 |
| | | | Standard Error of the Mean | 0.47 | 0.036 |

Table 3: Mean, standard deviation, and standard error of the mean derived from our measurements of WDS 09168-3407 B 1118.

## 4.0 Discussion

We present the results of our study of WDS 09168-3407 B 1118, a double star system, with the new position angle and separation measured at 341.4° and 3.36", respectively. The previous values derived in 1938 were 340.2° and 3.43". We do not trust the results from the most recent studies of the system in 2010, which yielded 325.9° and 6.88". This is because they are significantly disparate from both our results and those of the previous studies.

Based on historical data and our new measurements, we illustrate the variation in the separation of the two stars in Table 4 and plot it in Figure 3.

Table 4: compilation of the double-star systems' historical data sourced from the Washington Double Star Catalog. (link to data)

| Year | PA | Sep |
|---|---|---|
| 1911.24 | 334.1 | 3.337 |
| 1914.29 | 341.8 | 3.78 |
| 1924.45 | 341.8 | 3.69 |
| 1928.83 | 340.2 | 3.57 |
| 1930.34 | 340.0 | 3.53 |
| 1938.31 | 340.2 | 3.43 |
| 2010.5 | 325.9 | 6.88 |
| 2024.1425 | 341.4 | 3.36 |

We plotted the historical data of the system from Table 4. The plot in Figs. 3 shows the data points according to the times of studies of the B 1118 system, following the color sequence: Red (R), Red (R), Orange (O), Orange (O), Yellow (Y), Yellow (Y), Green (G), and Blue (B). The blue data point represents our measurements.

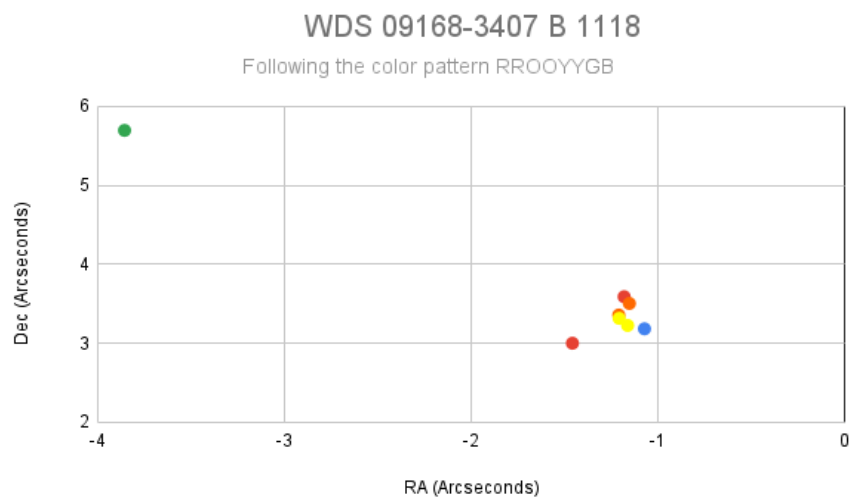

Figure 4: Graph illustrating the historical data alongside the recent measurements of WDS 09168-3407 B 1118.

Table 4 has information about the B 1118 system from when it was first observed in 1911 up to our latest data in 2024. A careful look at the separation column in Table 4 shows how far apart the star systems are over time, we see that the separation increased until 1914, and then it started decreasing. This strange change is mostly because of things like atmospheric effects, data errors, and uncertainties in measurements. Figure 3 shows this trend in colors: Red (R), Red (R), Orange (O), Orange (O),

Yellow (Y), Yellow (Y), Green (G), and Blue (B). So, based on this pattern, we think the separation will probably keep decreasing in the future.

Another important feature of the B 1118 double star system is the parallax values of the primary and secondary stars, which are 1.6725 mas and 1.6461 mas, respectively. These values imply that the systems are positioned at approximately the same distance from Earth, suggesting their closeness to each other and the possibility of gravitational binding.

To support this claim, examining the proper motions (in RA and Dec) of the system, as presented in Table 1, and utilized to calculate the ratio of proper motion (rPM) metric, resulting in a value of 0.033 (ultimately less than 0.2). This suggests that the system is a Common Proper Motion (CPM) pair. According to Harshaw (2016), CPM pairs typically indicate binary systems.

However, our plot in Figure 3 does not display any curvature or specific pattern, and Table 4 reveals that the system's separation once increased, reached a turning point, and is now decreasing. When considering the data from rPM and parallax, it leads us to the conclusion that the system may indeed be a binary star.

**5.0 Conclusion**

In this study, we measured the position angle and separation of the double star system WDS 09168-3407 B 1118 and compared them with historical data. Additionally, we calculated the parallax and proper motion of the primary and secondary stars to understand the nature of the system. Here are our main findings:

The separation of the system has consistently decreased over the past century, indicating probable gravitational binding with periods likely exceeding 100 years.

The parallax values for the primary and secondary stars, 1.6725 and 1.6461, respectively, suggest that they are at a similar distance from Earth and may be gravitationally bound.

The ratio of proper motion (rPM) metric for the system is 0.033, which is below 0.2, implying that the system is a common proper motion (CPM) pair.

Although the system's plot shows no curvature, considering the findings mentioned above, the system may still be a binary system. We recommend continuous observation to determine when the next turning point occurs, depicting the minimum separation.

**Acknowledgments**

This research was made possible by the Washington Double Star catalog maintained by the U.S. Naval Observatory. We would like to express our appreciation for the resources provided by the StelleDoppie catalog, managed by Gianluca Sordiglioni, as well as the assistance from Astrometry.net and the AstroImageJ software developed by Karen Collins and John Kielkopf.

Additionally, we acknowledge the European Space Agency (ESA) mission Gaia (https://www.cosmos.esa.int/gaia), and we are grateful for the data processed by the Gaia Data Processing and Analysis Consortium (DPAC, https://www.cosmos.esa.int/web/gaia/dpac/consortium). Funding for DPAC has been generously provided by national institutions, especially those participating in the Gaia Multilateral Agreement.

Our observations benefited from the 0.4m telescopes of the Las Cumbres Observatory Global Telescope Network at the Cerro Tololo facility in Chile.

Special thanks to the Pan-African Citizen Science e-Lab (PACS e-Lab) management for facilitating this research opportunity in Africa. We also extend our gratitude to Gianluca Sordiglioni for maintaining the informative Stelle Doppie site, Dr. Rachel Freed for the recorded research guide, and Kalee Tock for creating the plotting instructions.